\documentclass[twocolumn]{aastex62}

 \usepackage{amsmath, amsthm, amssymb,amsfonts}
\usepackage{graphicx}
\usepackage{dcolumn}
\usepackage{bm}
\usepackage{color}
\usepackage{multirow}

\submitjournal{ApJ}
\date{\today}
\shorttitle{Magnetic field kinks and folds in the solar wind}
\shortauthors{Tenerani et al.}

\begin{document}

\title{MAGNETIC FIELD KINKS AND FOLDS IN THE SOLAR WIND}

\author[0000-0003-2880-6084]{Anna Tenerani}
\affiliation{Department of Physics, University of Texas at Austin, TX 78712, USA}
  \email{Anna.Tenerani@austin.utexas.edu}
\author[0000-0002-2381-3106]{Marco Velli}%
 \affiliation{Department of Earth, Planetary, and Space Sciences, UCLA, Los Angeles, CA, 90095, USA}
 \author[0000-0002-6276-7771]{Lorenzo Matteini}
 \affiliation{Imperial College London, South Kensington Campus, London SW7 2AZ, UK}
\author[0000-0002-2916-3837]{Victor Réville}
\affiliation{Department of Earth, Planetary, and Space Sciences, UCLA, Los Angeles, CA, 90095, USA}
\affiliation{IRAP, Universit\'e Toulouse III - Paul Sabatier,
CNRS, CNES, Toulouse, France}
\author{Chen Shi}%
 \affiliation{Department of Earth, Planetary, and Space Sciences, UCLA, Los Angeles, CA, 90095, USA}

\author[0000-0002-1989-3596]{Stuart D. Bale}
\affil{Physics Department, University of California, Berkeley, CA 94720-7300, USA}
\affil{Space Sciences Laboratory, University of California, Berkeley, CA 94720-7450, USA}
\affil{The Blackett Laboratory, Imperial College London, London, SW7 2AZ, UK}
\affil{School of Physics and Astronomy, Queen Mary University of London, London E1 4NS, UK}

\author{Justin Kasper}
\affil{University of Michigan, Ann Arbor, MI, USA}

\author[0000-0002-0675-7907]{J. W. Bonnell}
\affil{Space Sciences Laboratory, University of California, Berkeley, CA 94720-7450, USA}

\author{Anthony W. Case}
\affiliation{Smithsonian Astrophysical Observatory, Cambridge, MA, USA}

\author[0000-0002-4401-0943]{Thierry {Dudok de Wit}}
\affil{LPC2E, CNRS and University of Orl\'eans, Orl\'eans, France}

\author[0000-0003-0420-3633]{Keith Goetz}
\affiliation{School of Physics and Astronomy, University of Minnesota, Minneapolis, MN 55455, USA}

\author[0000-0002-6938-0166]{Peter R. Harvey}
\affil{Space Sciences Laboratory, University of California, Berkeley, CA 94720-7450, USA}

\author{Kristopher G. Klein}
\affil{Lunar and Planetary Laboratory and Department of Planetary Sciences University of Arizona
Tucson, AZ 85719, USA}

\author{Kelly Korreck}
\affil{Smithsonian Astrophysical Observatory, Cambridge, MA, USA}

\author{Davin Larson}
\affiliation{Space Sciences Laboratory, University of California, Berkeley, CA 94720-7450, USA}

\author{Roberto Livi}
\affiliation{Space Sciences Laboratory, University of California, Berkeley, CA 94720-7450, USA}

\author[0000-0003-3112-4201]{Robert J. MacDowall}
\affil{Solar System Exploration Division, NASA/Goddard Space Flight Center, Greenbelt, MD, 20771}

\author[0000-0003-1191-1558]{David M. Malaspina}
\affil{Laboratory for Atmospheric and Space Physics, University of Colorado, Boulder, CO 80303, USA}

\author[0000-0002-1573-7457]{Marc Pulupa}
\affil{Space Sciences Laboratory, University of California, Berkeley, CA 94720-7450, USA}

\author{Michael Stevens}
\affiliation{Smithsonian Astrophysical Observatory, Cambridge, MA 02138 USA}

\author{Phyllis Whittlesey}
\affiliation{Space Sciences Laboratory, University of California, Berkeley, CA 94720-7450, USA}

\begin{abstract}
Parker Solar Probe (PSP) observations during its first encounter at 35.7 $R_\odot$ have shown the presence of magnetic field lines which are strongly perturbed to the point that they produce local inversions of the radial magnetic field, known as switchbacks. Their  counterparts in the solar wind velocity field are local enhancements in the radial speed, or jets, displaying (in all components) the velocity-magnetic field correlation typical of large amplitude Alfv\'en waves propagating away from the Sun.  Switchbacks and radial jets have previously been observed over a wide range of heliocentric distances by Helios, WIND and Ulysses, although they were prevalent in significantly faster streams than seen at PSP. Here we study via numerical MHD simulations the evolution of such large amplitude Alfv\'enic fluctuations by including, in agreement with observations, both a radial magnetic field inversion and an initially constant total magnetic pressure.  Despite the extremely large excursion of magnetic and velocity  fields, switchbacks are seen to persist for up to hundreds of Alfv\'en crossing times before eventually decaying due to the parametric decay instability. Our results suggest that such switchback/jet configurations might indeed originate in the lower corona and survive out to PSP distances, provided the background solar wind is sufficiently calm, in the sense of not being pervaded by strong density fluctuations or other gradients, such as stream or magnetic field shears, that might destabilize or destroy them over shorter timescales.

\end{abstract}


\section{Introduction} 
Measurements in the solar wind have revealed that the relatively steady wind emanating from coronal holes, typically fast wind streams with speed around and above $700$~km/s, occasionally display local magnetic field polarity inversions known as ``switchbacks'', embedded within the continuous flux of Alfv\'enic turbulence emanating from the sun.  Switchbacks  have been observed over a wide range of heliocentric distances ($R$), from $R\simeq0.3$~AU~\citep{horbury_MNRAS} all the way out to $R\geq 1.3$~AU~\citep{balogh_1999,marcia_2013}, and now also by the Parker Solar Probe~(PSP).  

While approaching perihelion at a radial distance of $R=35.7\,R_\odot$ (solar radii), PSP sampled solar wind streams with speed $V$ ranging between $V\simeq 300-500$~km/s. Although the wind speed remains around values usually attributed to the slow wind (a highly variable wind with compressible and incoherent turbulent fluctuations) these new observations surprisingly revealed the  occurrence of large amplitude Alfv\'enic fluctuations much like what is found in the fast wind, and even more prominently so. Wind streams indeed appear to be pervaded by sequences of extremely large fluctuations in the radial magnetic field, with amplitudes of the same order of the background magnetic field, and switchbacks often leading to an almost complete backward rotation of the magnetic field itself  (see for instance~\citet{horbury_psp, case_psp}, this issue). 

Early studies have shown that the correlation of velocity and magnetic field fluctuations during switchbacks corresponds to that of Alfv\'enic fluctuations propagating away from the Sun. That the radial field inversions are in fact highly kinked and folded magnetic field lines,  rather than magnetic field lines connecting back to the Sun, was suggested by~\citep{balogh_1999} and was demonstrated most clearly by analysis of the proton-alpha differential speed in the fast streams, where the alpha particles, always moving faster than protons, were seen to flip behind during switchbacks~\citep{yamauchi_2004}. The electron strahl was also shown to remain aligned with the magnetic field, while the core and beam of the proton distributions also flipped~\citep{marcia_2013}.
 Such folds/kinks of the magnetic field, despite their coherent appearance, are an intrinsic part of the Alfv\'enic turbulent spectrum and can occur over a wide range of scales, from seconds to several minutes~\citep{horbury_MNRAS} or tens of minutes~\citep{marcia_2013}, but smaller than microstream scales, that dominate the lower frequency part of the radial energy spectrum at periods of the order of a couple of days~\citep{marcia_1995}.   \citet{matteini_2014} showed that switchbacks  are the magnetic manifestation of impulsive enhancements that can be observed in the  radial  bulk speed of the wind, the so-called radial jets~\citep{gosling_2009}, or spikes. The connection between  radial jets and switchbacks follows directly from the velocity-magnetic field correlation that characterizes Alfv\'en waves, complemented by the fact that the total magnetic field magnitude remains nearly constant allover the fluctuations---the latter being a well established, observed property of Alfv\'enic turbulence in the solar wind (e.g., \citet{barnes_1974,tsuru,matteini_2014}.) 

Whether switchbacks are the remnant of some process in the corona related to wind formation and acceleration  remains an open question. Remote observations bring evidence of a plethora of transient and impulsive events such as plumes, spicules and rays~\citep{rouafi_2016}, but the causal connection between those events observed remotely and  in-situ measurements of the wind has not  been established as of yet. On the other hand,  switchbacks lasting from a few minutes to a couple of hours seen by Ulysses appear to be consistent with network-scale size structures at the sun's surface, and it has been suggested  that they may originate in the low corona from reconnection between emerging closed magnetic flux from the sun's surface and pre-existing open magnetic fields~\citep{yamauchi_2004}.  

If switchbacks were the byproduct of reconnection as suggested, then the problem is that of maintaining the highly kinked field lines all the way out to tens of solar radii where they are currently observed. Indeed, a highly kinked Alfv\'enic fluctuation is not expected to survive beyond a few of solar radii, because at the low-$\beta$ (thermal to magnetic pressure ratio) values typical of the solar corona, strong coupling with compressible modes naturally tend to unfold and stretch magnetic field lines until the dominant polarity is recovered. Previous work by~\citet{Landi_2005}, who considered a kinked Alfv\'en wave not in pressure equilibrium with the surroundings, indeed ruled out the possibility of  a coronal origin of switchbacks for those reasons. 

Here, motivated by the new observations of the Parker Solar Probe, we show  that a highly kinked Alfv\'enic wave packet characterized by constant total magnetic field strength  can persist for up to hundreds of Alfv\'en crossing times, suggesting that such fluctuations may indeed originate in the corona and propagate all the way out to PSP distances, provided the wind in not pervaded by strong and frequent fluctuations in plasma quantities which may destroy the switchback  on shorter timescales.
 \begin{figure}[htb]
\includegraphics[width=0.5\textwidth]{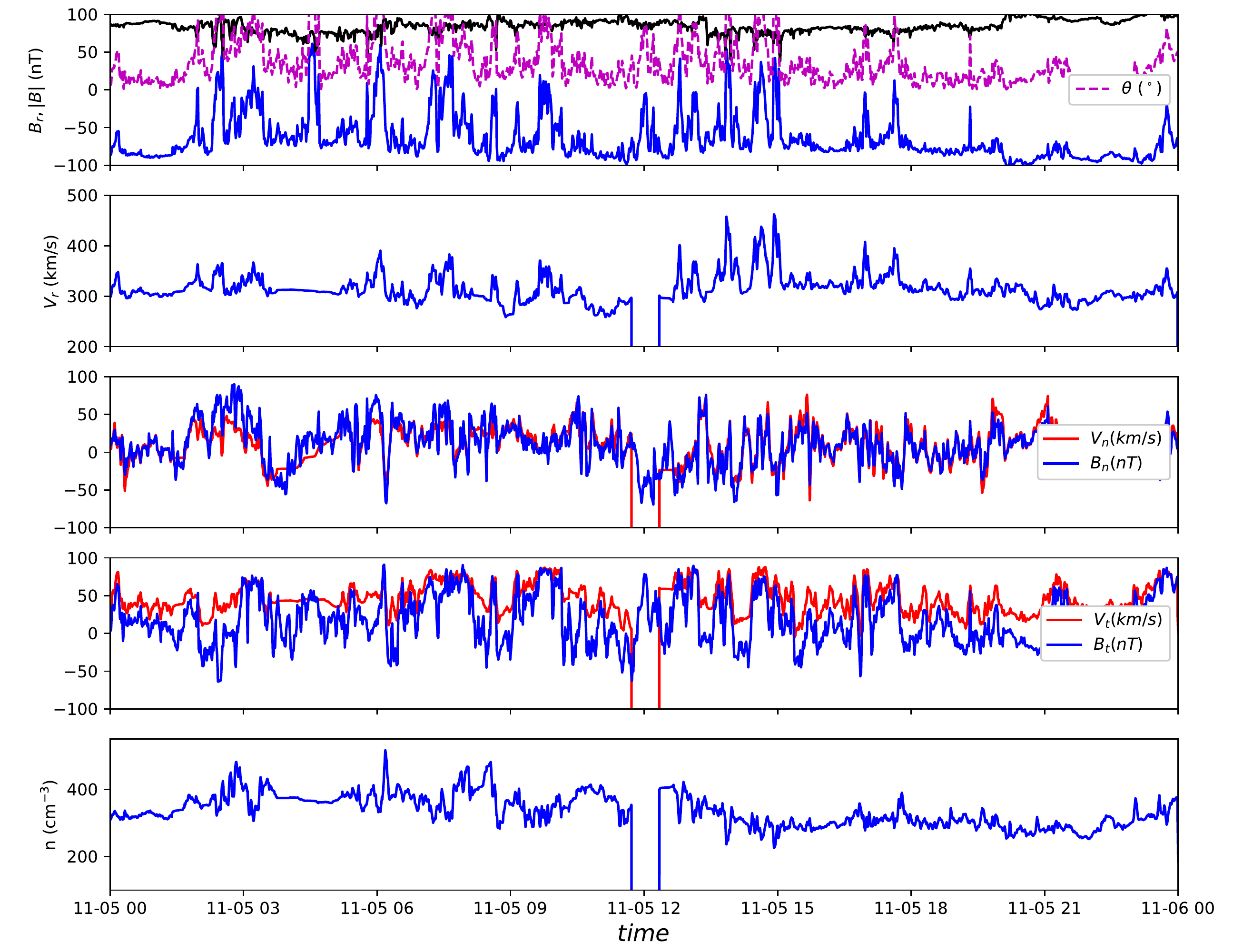}
\caption{Data sample of  magnetic field and plasma measurements gathered  during the first perihelion  by Parker Solar Probe, showing a sequence of switchbacks and radial jets. The upper plot displays the radial component of the magnetic field  (blue), the magnitude of the total magnetic field (black) and the angle between the radial and the total magnetic field (magenta). The second panel shows the radial proton speed, and the third and fourth panel the transverse components (namely, the normal and tangential components in RTN coordinates) of the magnetic and velocity fields (blue and red colors, respectively). The last panel shows the proton number density.}
\label{psp}
\end{figure}

\section{Data sample}
 Fig.~\ref{psp} shows a data sample of  magnetic field and plasma measurements gathered  during the first perihelion by the FIELDS~\citep{bale_2016} and SWEAP~\citep{kasper_2016}  instrument suites. Switchbacks clearly stand out as huge spikes in the radial magnetic field and  velocity profiles ($B_r$ and $V_r$) lasting about 20 minutes in this particular interval, shown in blue color in the first and second panels. The radial magnetic field changes from roughly  a base value of $B_r\simeq-70$~nT up to $B_r\simeq+60$~nT, whereas the radial speed changes from a base value of about $V\simeq300$~km/s to $V\simeq390$~km/s  and in some cases even up to $V\simeq450$~km/s. The magenta color in the first panel represents the angle $\theta$ between the radial  and the total magnetic field: while in the absence of strong fluctuations the magnetic field is aligned with the radial ($\theta=0$), in correspondence of the switchbacks it turns to more than $\theta=100^\circ$, indicating an almost complete rotation of the magnetic field in the opposite direction (notice that to compute the angle we have taken the reversed sign of $B_r$, which is directed radially inwards in the absence of fluctuations). At the same time, the total magnetic field magnitude $|{\bf B}|=B$, shown in black color in the first panel, remains relatively constant during such rotations. This is a common condition found in the solar wind implying that such  Alfv\'enic fluctuations are bound to rotate on a sphere of constant radius of magnitude $B$. In other words, Alfv\'enic fluctuations in the solar wind are spherically polarized~\citep{tsuru,matteini_2014}.  The  normal and tangential components of the magnetic (blue) and velocity (red) fields are shown in the third and fourth panels, respectively, displaying the typical magnetic-velocity field correlation  of Alfv\'en  waves propagating outwards from the sun, while  the proton number density is shown in the last panel.  As can be seen, occasionally, both the density and the magnitude of ${\bf B}$ display some level of fluctuations, and it will be of interest to investigate in more detail in the future the nature and role of such compressible effects.  
 
 Inspired by these and  other similar observations, in the next sections  we first provide  a  model for switchbacks and then we investigate via numerical simulations their evolution. 

\section{Analytical model for ``switchbacks" and numerical setup}

 We take a highly kinked Alfv\'en wave packet propagating in a homogeneous plasma with density $\rho_0$ and pressure $p_0$  along the in-plane background magnetic field ${\bf B}_0$, that we take along the ${ \hat y}$ direction. It has been known for a long time that unidirectional Alfv\'enic fluctuations provide an exact solution  to the  nonlinear, compressible Magnetohydrodynamic (MHD) system if the strength of the total  magnetic field is also constant~\citep{barnes_1974}.  Three dimensional Alfv\'enic solutions satisfying the constraint $B^2=const$ can be obtained numerically~\citep{valentini_2019}, however, a solution in closed analytical form has yet to be found to model Alfv\'enic wave packets of arbitrary amplitude. Therefore, here we  will adopt a reduced  model in two dimensions. 
 
 In order to construct our Alfv\'enic wave packet it is useful to introduce the two-dimensional magnetic scalar potential $\psi(x,y)$ and define the  magnetic field as 
\begin{equation}
{\bf B}=  {\boldsymbol\nabla}\times\psi(x,y){\hat z}+B_z(x,y){\hat z} + B_0{\hat y}. 
\label{potential}
\end{equation}
Any exact Alfv\'enic solution can be constructed by starting from eq.~(\ref{potential}) and by imposing  the constraint $B^2=const$ to determine the  out-of-plane component of the magnetic field, 
\begin{equation}
B_z(x,y)^2 = B^2-(B_x(x,y)^2+B_y(x,y)^2), 
\end{equation}
whereas the velocity fluctuation ${\bf u}$ follows directly from the Alfv\'enicity condition, ${\bf u}=-{\bf \delta B}/\sqrt{4\pi \rho_0}$, where ${\bf \delta B}$ is the fluctuating magnetic field with respect to its average value.  Starting from the profile considered by \citet{Landi_2005}, we model a local magnetic polarity inversion (the switchback) by specifying the following functional form for $\psi(x,y)$, 
\begin{equation}
\psi(x,y)= -\psi_0 \left( e^{-r_1^2}-e^{-r_2^2} \right),
\end{equation}
where 
\begin{equation}
r_{1,2}^2=\left(\frac{x-x_{1,2}}{\ell_x}\right)^2+\left(\frac{y-y_{1,2}}{\ell_y}\right)^2.
\end{equation}
  \begin{figure}[tb]
 \includegraphics[width=0.4\textwidth]{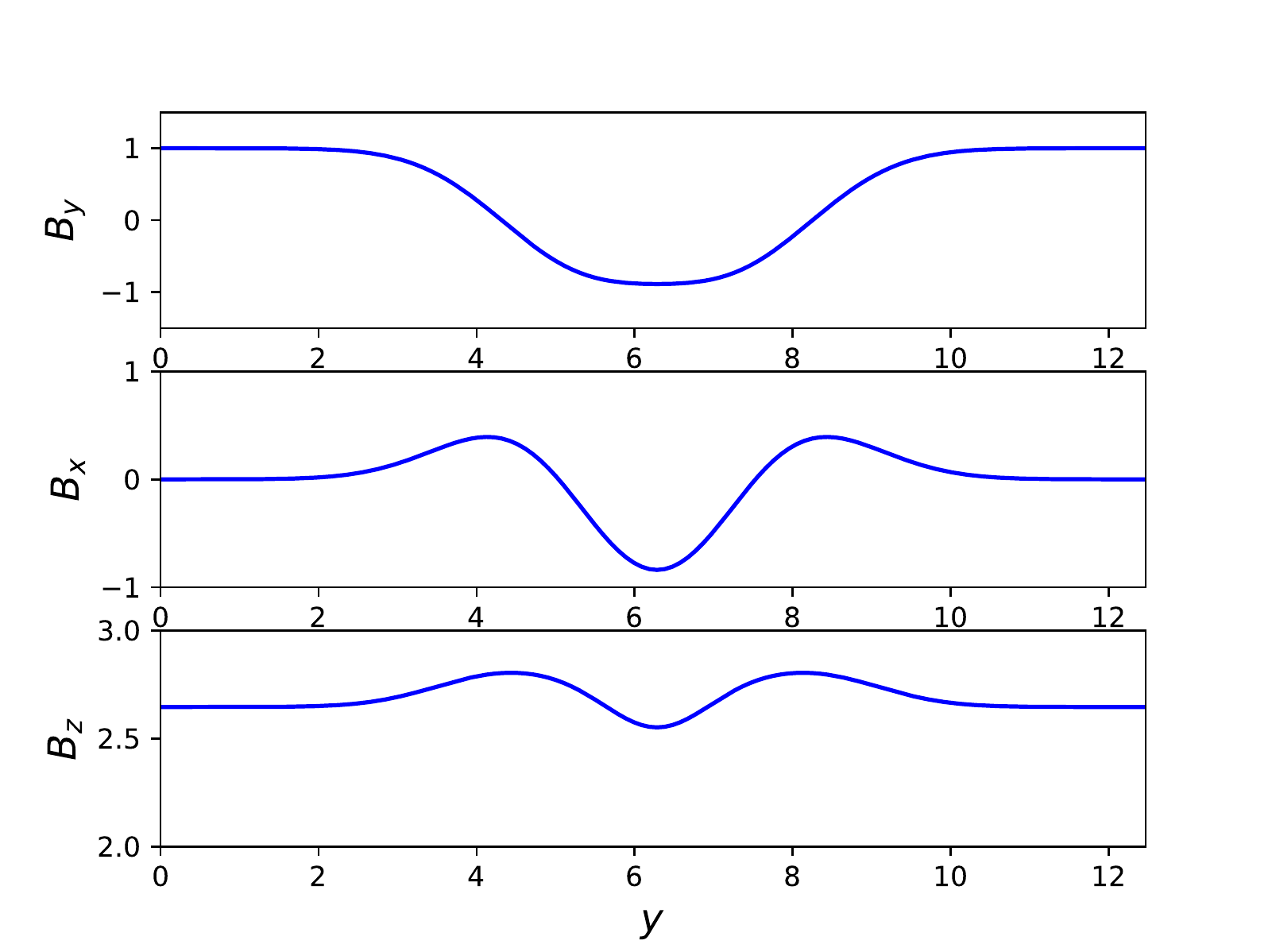}\\
 \includegraphics[width=0.4\textwidth]{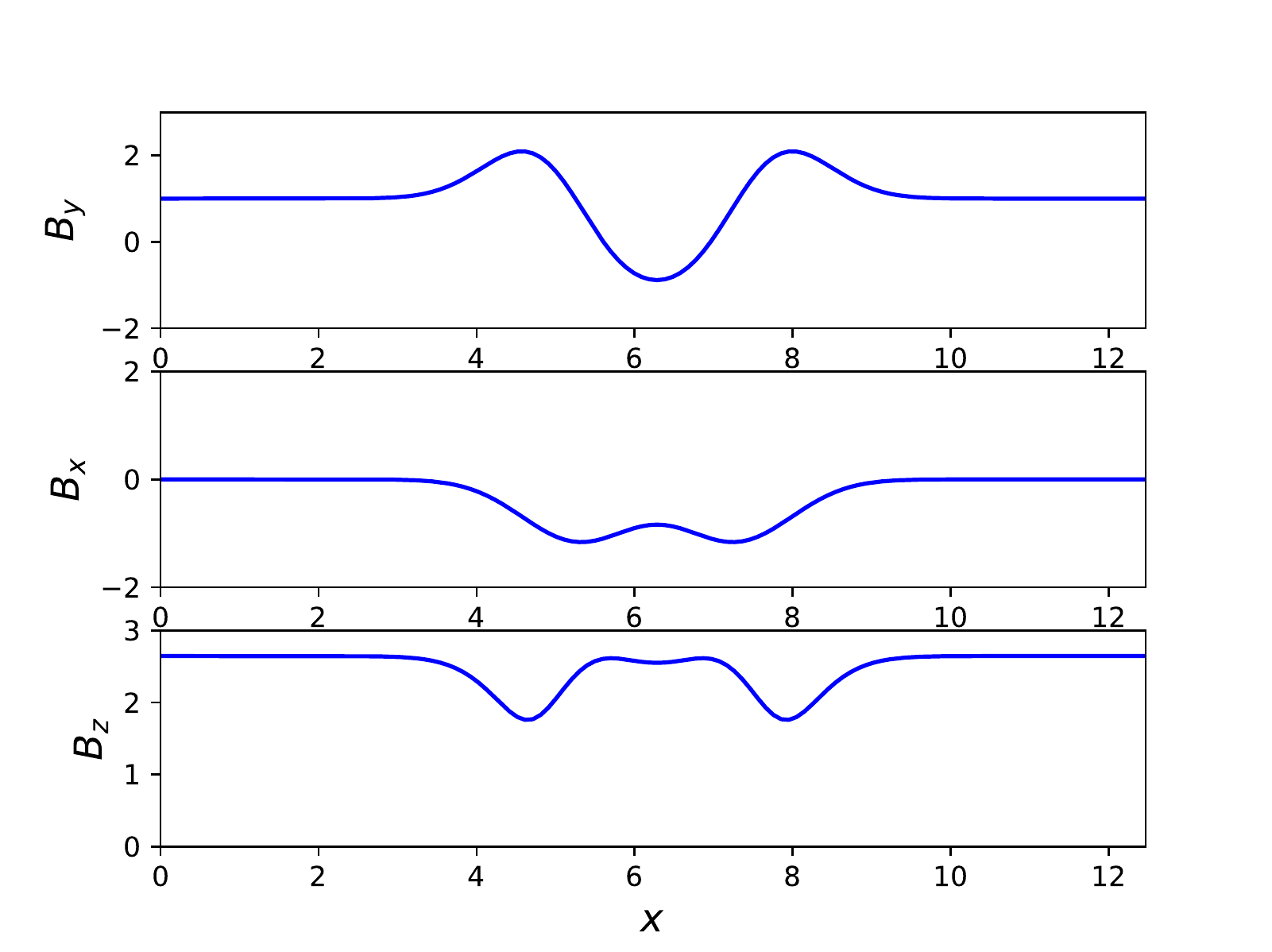}\\
 \includegraphics[width=0.4\textwidth]{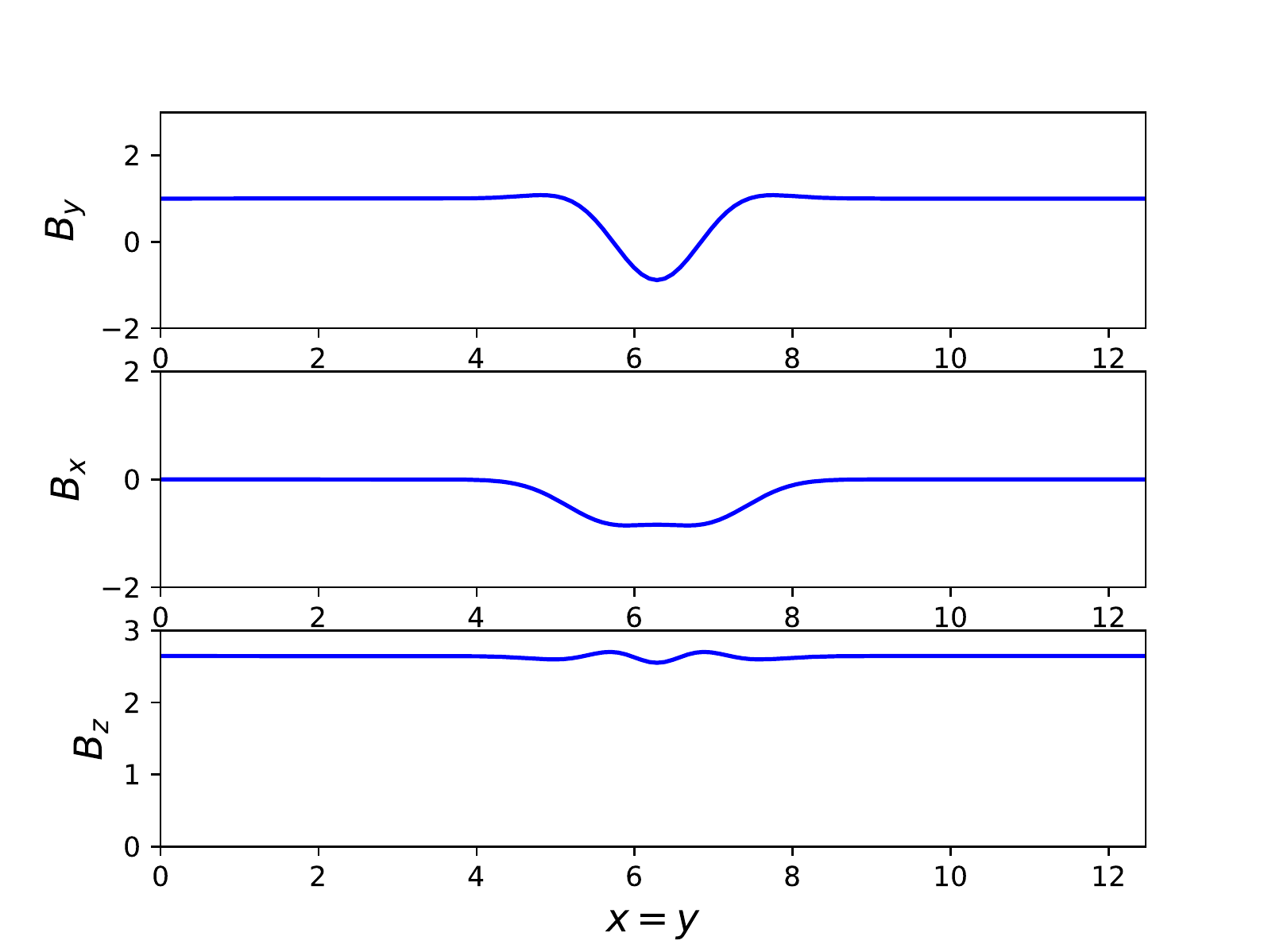}
\caption{Cuts of the magnetic field used as initial configuration: along the switchback at $x=L_x/2$ (upper panel); across the switchback, at $y=L_y/2$ (middle panel); at an oblique angle of $45^\circ$ (bottom panel).}
\label{cuts}
\end{figure}

\begin{figure*}[tb]
 \includegraphics[width=0.5\textwidth]{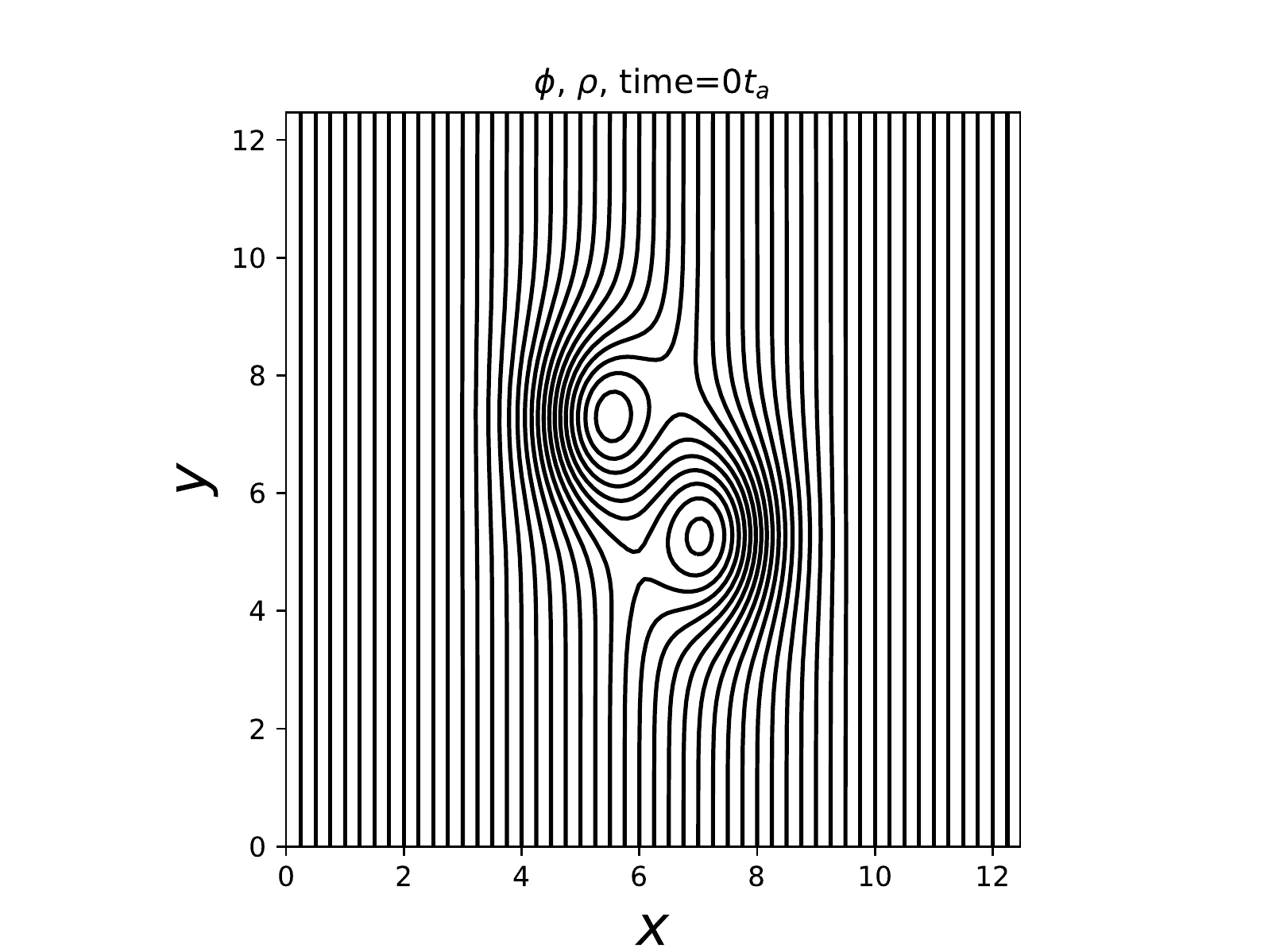}\includegraphics[width=0.5\textwidth]{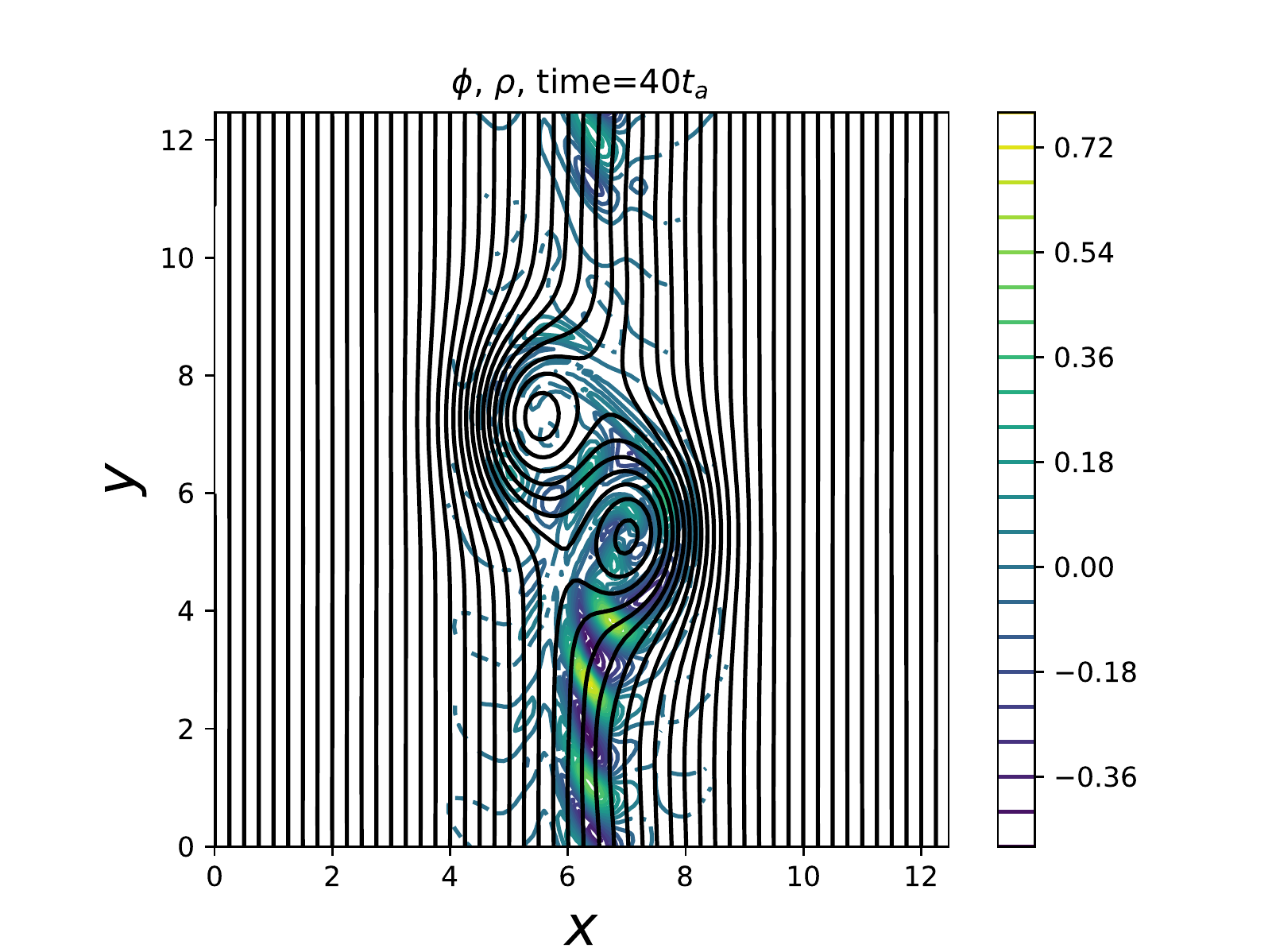}\\
 \includegraphics[width=0.5\textwidth]{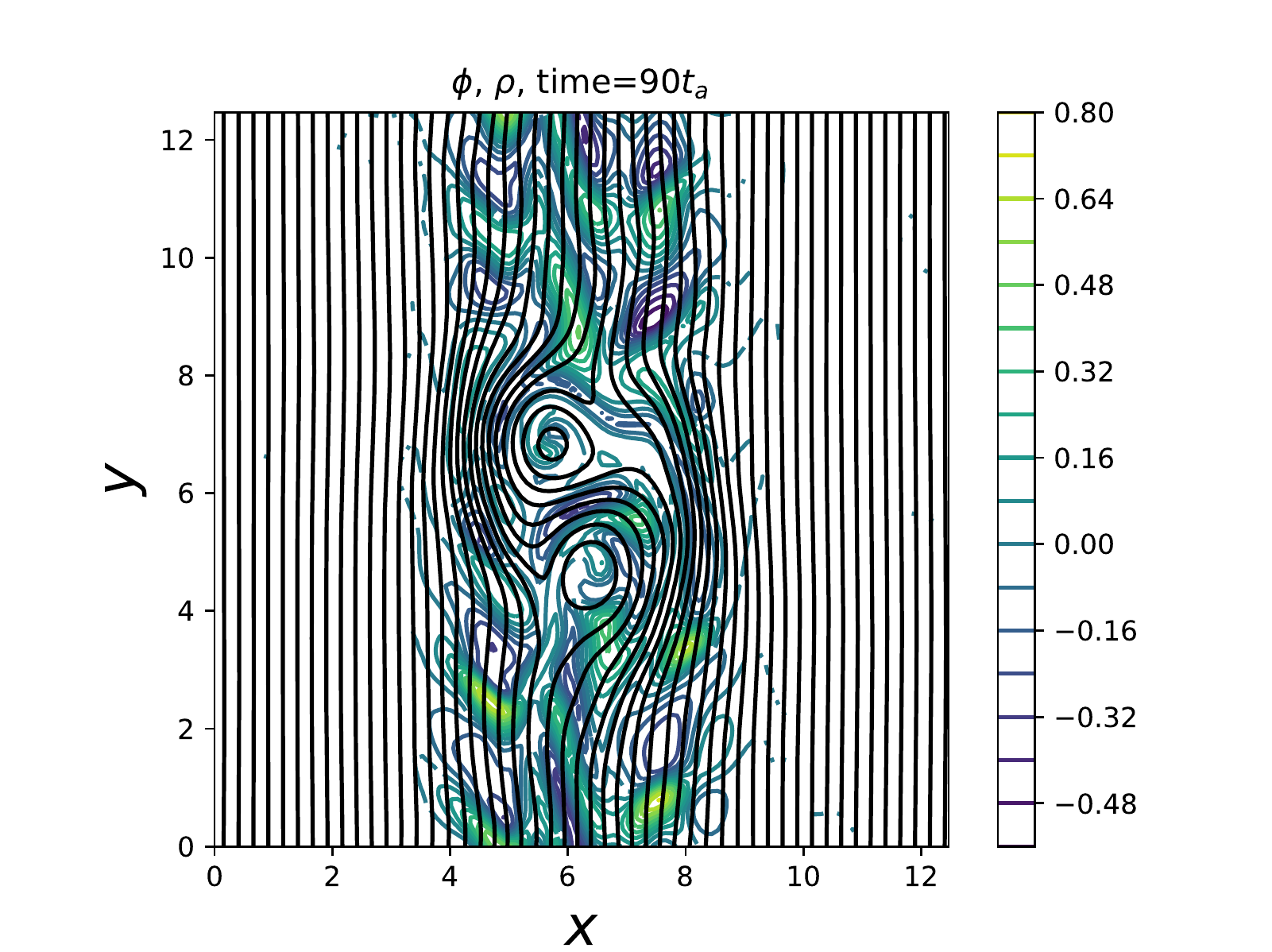}\includegraphics[width=0.5\textwidth]{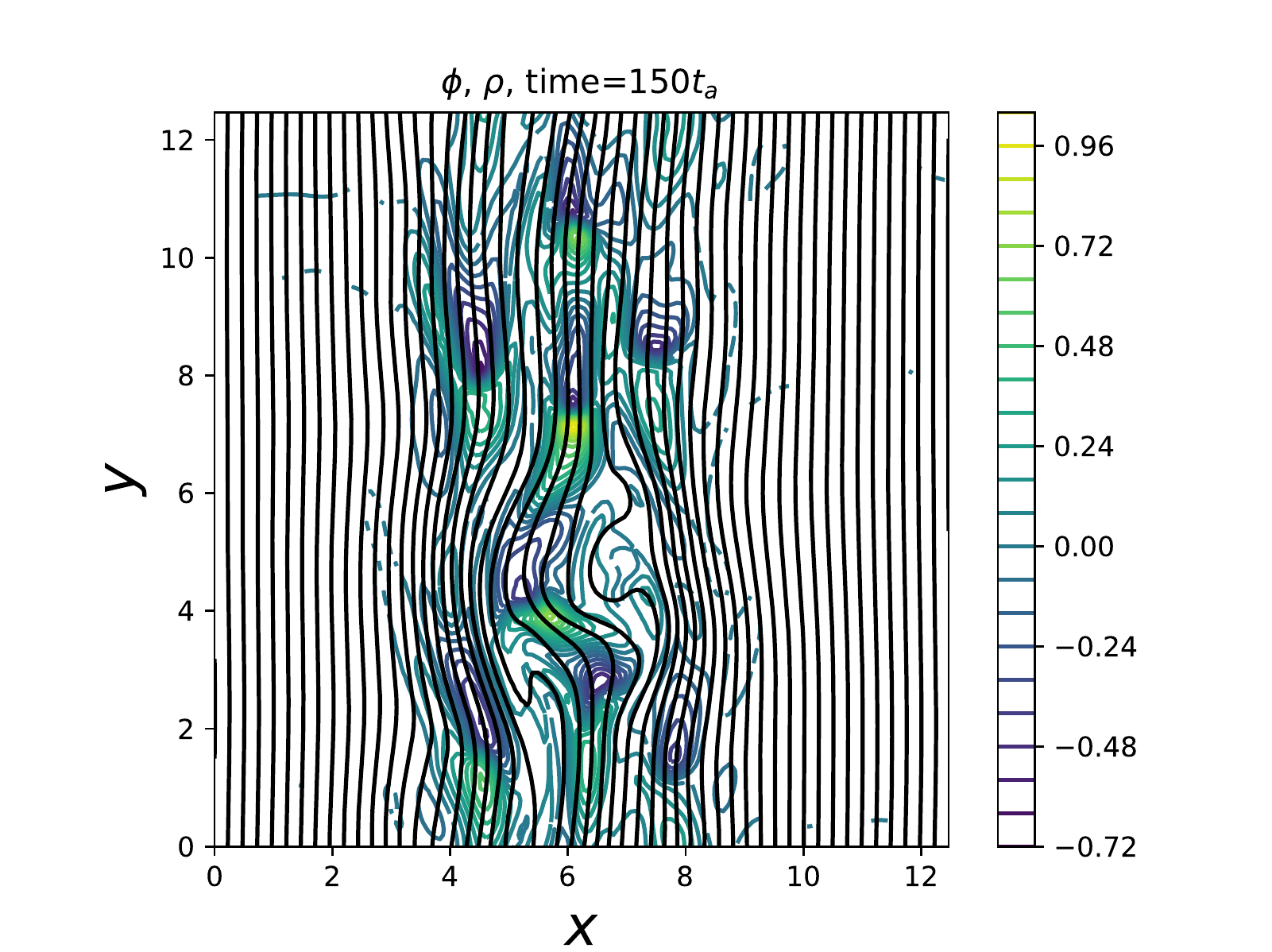}
\caption{Magnetic field lines (black) and contours of density fluctuations (color coded)  from $t=0$ until $t=150\,t_a$ (run 1). Only contour values corresponding to $|(\rho-\rho_0)/\rho_0|\geq10^{-3}$ are displayed.}
\label{evol}
\end{figure*}

In this work we  use a  2.5D (two spatial coordinates and three dimensional fields) compressible MHD code where  an adiabatic closure is assumed. The code is periodic in both the $y$ and the $x$-direction.  For our simulations we choose a case study corresponding to an anisotropic wave packet with $\ell_x=1$, $\ell_y=1.5$, $|x_1-x_2|=|y_1-y_2|=2$, and amplitude $\psi_0=2$. These parameters ensure the presence of an inversion of the magnetic field with minimum value of $B_y/B_0=-0.9$  at the center of the simulation domain. In Fig.~\ref{cuts} we show three cuts of the initial configuration along the switchback,  at $x=L_x/2$ (upper panel), across the switchback, at $y=L_y/2$ (middle panel) and at an angle of $45^\circ$ (bottom panel). The resulting in-plane magnetic field lines can be seen in the upper left panel of Fig.~\ref{evol}.  Notice that because of the limitations imposed by the 2D geometry, the out-of-plane component of the magnetic field is of the same order of magnitude of the guiding  magnetic field ${\bf B}_0$, so that in fact the switchback propagates at an angle $\theta\simeq69^\circ$ with respect to the total average magnetic field. However, this geometrical limitation does not compromise the main goal of this study, which is  to investigate the stability of folds/kinks of the magnetic field along the propagation direction. The plasma beta ($\beta=8\pi p_0/B_0^2$) is set to $\beta=0.1$ in most cases, with one simulation having $\beta=1$. We keep a constant mesh resolution $\Delta x=\Delta y=0.1$ and we perform several simulations with different  box length  $L_y$ (along the propagation direction), while $L_x=2\times 2\pi$. A list of the variable parameters for each simulation run can be found in Table~\ref{sims_param}. Finally, in order to facilitate the numerical analysis, we perform a Galilean transformation and integrate the set of MHD equations in the co-moving frame of the Afv\'en wave packet. 
\begin{table}[htp]
\caption{Simulation parameters.}
\begin{center}
\begin{tabular}{l c c c}
\toprule
run \#&$L_y$&$\beta$	\\
run 1 &$2\times2\pi$	& \multirow{6}{*}{0.1}	\\
run 2 &$4\times2\pi$	 &\\
run 3 &$6\times2\pi$	\\
run 4 &$8\times2\pi$	\\
run 5 &$16\times2\pi$\\
run 6 &$32\times2\pi$\\
run 7 &$2\times2\pi$	&1\\
\toprule
\end{tabular}
\end{center}
\label{sims_param}
\end{table}%

\section{Results}
The simulations considered in this work display a similar qualitative behavior. We  therefore begin this section  by focusing on run 1, which will serve as a reference to give an overview of, and discuss, our main results.  In Fig.~\ref{evol} we show the time evolution of the switchback, where contours of density fluctuations (color coded)  are superposed to magnetic field lines (black lines) from $t=0$ until  the magnetic field  lines have become almost aligned with the dominant magnetic field polarity. The switchback remains stable for several tens of Alfv\'en crossing times $t_a=\ell_y/v_a$, where $v_a=B_0/\sqrt{4\pi\rho_0}$. Eventually, starting from about $t\simeq 60\, t_a$, it begins to interact with density fluctuations via a process that displays features consistent with the parametric decay instability~\citep{galeev_sov_phys_1963,derby}, a mechanism known to cause growing density fluctuations and wave reflection. 

Interaction with density fluctuations and partial reflection indeed cause the initial switchback to ultimately ``unfold" and disrupt until the magnetic field lines have turned to the dominant polarity.  The complete disruption of the switchback,  however, appears to be a relatively slow process. As can be seen in the bottom right panel of Fig.~\ref{evol}, where we show the evolved switchback at time $t=150\,t_a$,  the central magnetic field lines have become aligned with ${\bf B}_0$, but there are other regions in the trailing edge of the wave-packet displaying significant distortion including local polarity inversions. The complete unfolding of the initial switchback is found to occur at a much later time, around $t\simeq 190\,t_a$. 

Parametric decay has been studied in great detail in many different conditions, from MHD~\citep{DelZanna_JGR_2001,malara_Phys_fluids_1996,anna1, reville,shoda_a} to hybrid kinetic models~\citep{vasquez,matteini_GRL_2010}. Those works have studied parametric decay of planar fluctuations both in the  case of  coherent, monochromatic large amplitude waves and broad-band fluctuations. More recently, also  non-planar fluctuations have been studied, although in a larger $\beta$ plasma~\citep{primavera_2019}. However, only Alfv\'enic fluctuations that fill space homogeneously  have been considered so far. The fact that in our case the wave packet is localized in space has implications on the resonance conditions underlying the decay process,  the interaction time $\tau_i$ between compressible fluctuations and the wave-packet being limited to  $\tau_i\simeq \ell_y/(|c_s-v_a|)$, where $c_s$ is the sound speed. In our periodic simulations the interactions will resume after the time interval $\Delta t$ required to travel a length $L_y$, $\Delta t\simeq L_y/(|c_s-v_a|)$. In this regard we performed several simulations to explore the effect of the box size on the evolution of our switchback.

\begin{figure}[tb]
 \includegraphics[width=0.4\textwidth]{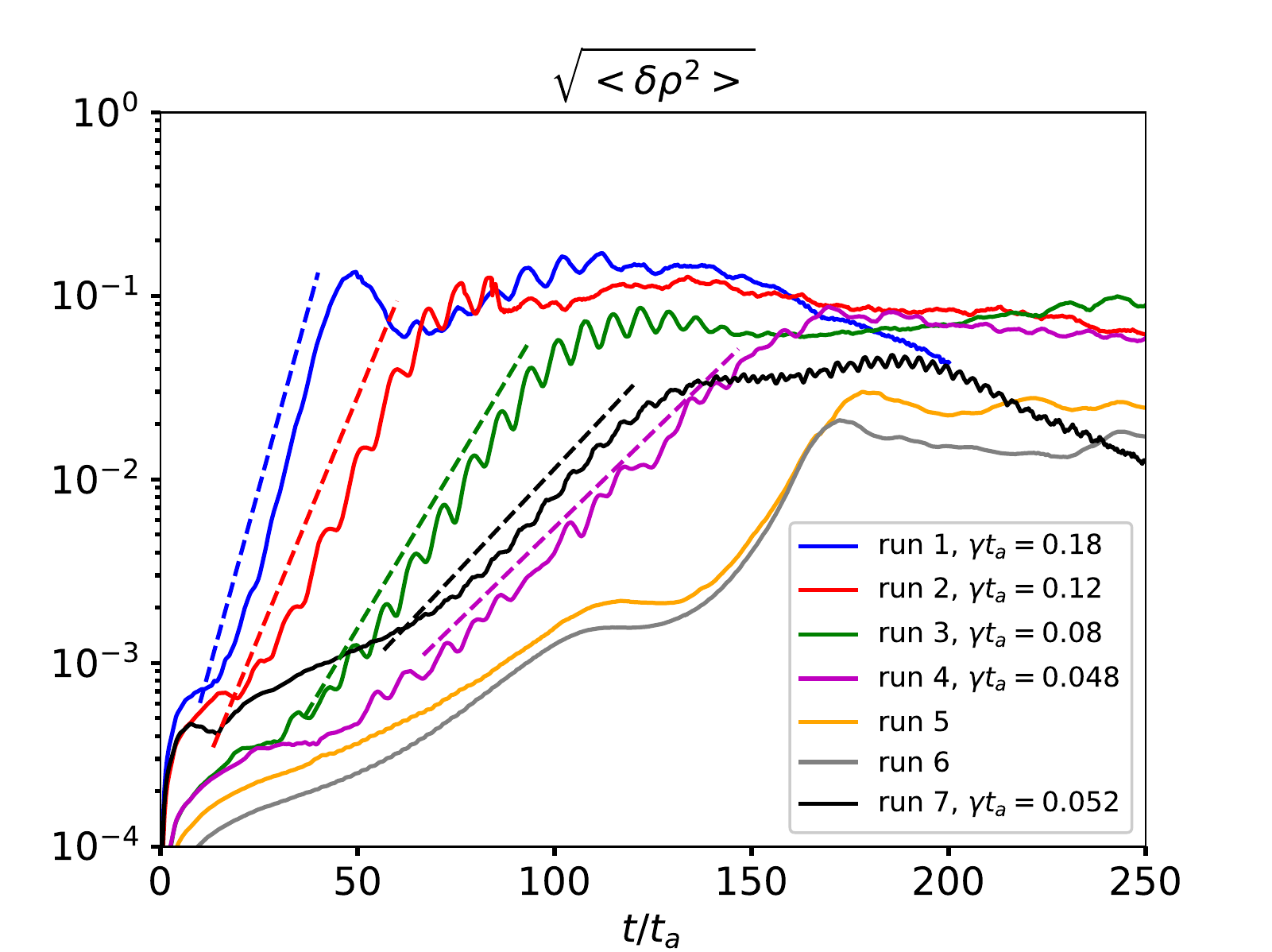}
\caption{Time evolution of the rms of density fluctuations. Estimated growth rates $\gamma$ are indicated in the legend for run 1--run~4 and run 7. Dashed lines represent the interval where  the growth rate for each simulation has been estimated.}
\label{gamma}
\end{figure}

In Fig.~\ref{gamma}  we show in linear-logarithmic scale the time evolution of the rms amplitude of the density fluctuations for each run. From run 1 to run 4 and in run 7 we can identify an initial stage during which density fluctuations grow exponentially until saturation. The growth rate $\gamma$, not so surprisingly,  is found to scale roughly as $\gamma t_a\sim \ell_y/L_y$, with the most unstable case having $\gamma t_a=0.18$.   By way of comparison, a monochromatic wave with relative amplitude $\delta B/B_0\sim 2$ and wavelength $\lambda=\ell_y$ has a maximum growth rate $\gamma t_{a}\simeq 3.7$ (that we estimated by solving numerically the linear dispersion relation. See, e.g., \citet{derby}), much larger than the growth rate of our localized wave-packet. The estimated growth rates for these runs are indicated in the legend of Fig.~\ref{gamma}, with the dashed lines showing the interval over which they have been estimated. After saturation, a second weaker growth  in the density rms can be observed, that we interpret as a secondary decay of  the reflected wave, as reported in previous work~\citep{DelZanna_JGR_2001}. Since parametric decay tends to be stabilized as  $\beta$ increases (at fixed amplitude),  we performed one simulation (run 7) with the same parameters as in run 1 but with $\beta=1$ and verified that  the growth rate is significantly reduced, confirming that the process leading to density growth is  a form of parametric instability. 

\begin{figure}[htb]
 \includegraphics[width=0.4\textwidth]{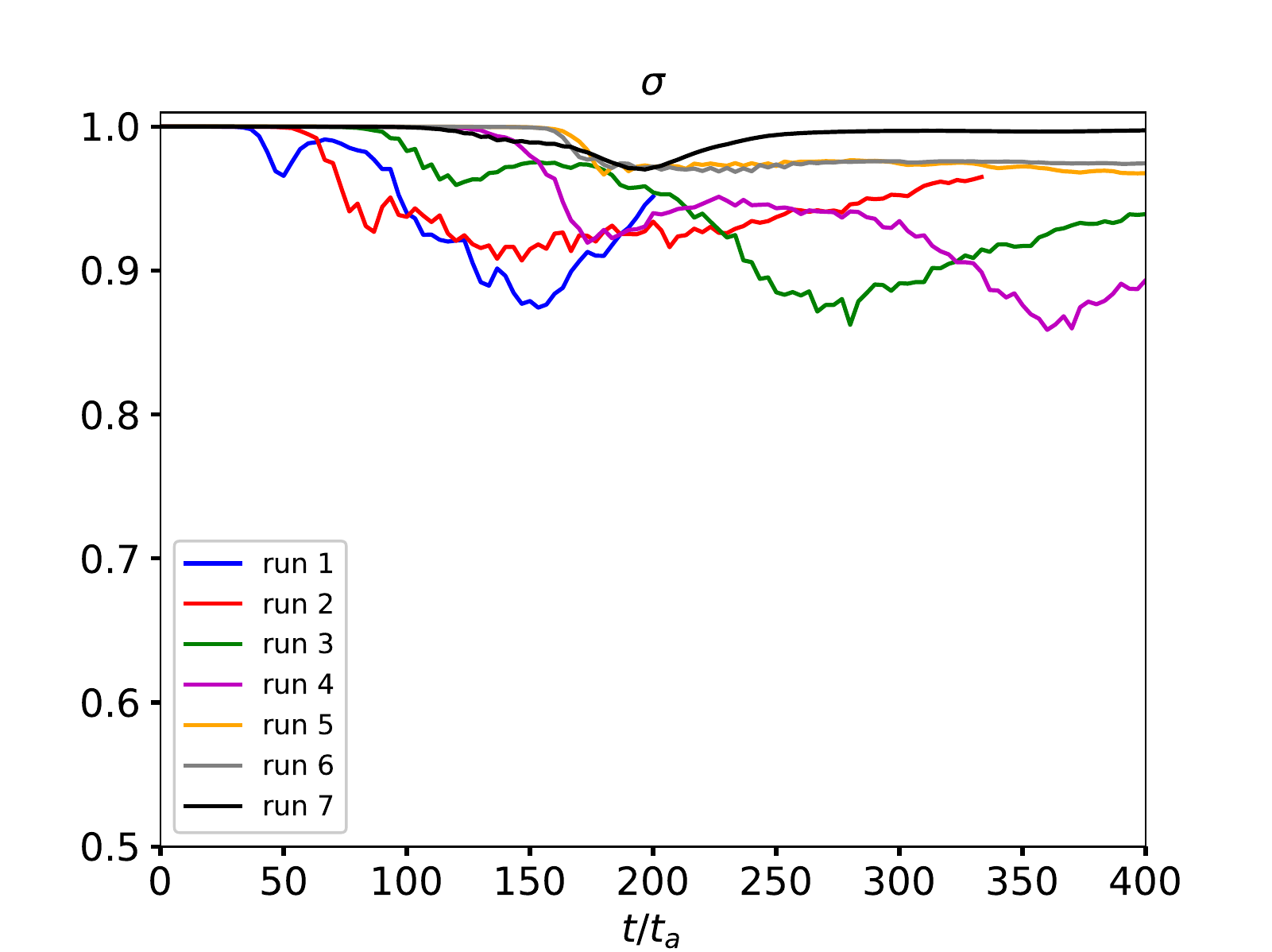}
\caption{Evolution of the average cross helicity~$\sigma$ for run~1-run~7.}
\label{sigma}
\end{figure}

Fig.~\ref{sigma}  displays the time evolution of the total average cross helicity ($\sigma$), that gives us a measure of the overall  Alfv\'enic correlation between velocity and magnetic fields: $\sigma=\pm1$ corresponds to an ensemble of perfectly correlated or anti-correlated velocity and magnetic field fluctuations  propagating forwards and backwards, respectively; $\sigma=0$ instead corresponds to an ensemble of balanced counter-propagating Alfv\'enic fluctuations. As can be seen,  the cross helicity drops when saturation of density fluctuations is attained, a feature typical of the nonlinear stage of  parametric decay, and it increases again following the trend of density fluctuations, consistent with the occurrence of a secondary decay of the (partially) reflected wave. However, the value of the cross-helicity remains significantly large and close to unity ($\sigma\gtrsim0.8$) in all of our runs,  implying that wave reflection is strongly inhibited in the case of our switchback, contrary to previous work that considered space-filling fluctuations and where cross helicity was found to drop all the way down to negative values (e.g., \citet{DelZanna_JGR_2001,anna1}).  

The trend of the growth rate to decrease with the system size is observed until the box length $L_y$ becomes sufficiently large, and then the evolution  becomes independent of $L_y$, at least for the duration of our simulations until   $t=466\,t_a$. Contrary to the previous cases, both run 5 and run 6, which correspond to the largest system sizes considered ($L_y/(2\pi)=16$ and $L_y/(2\pi)=32$, respectively),  display roughly two stages: an initial stage lasting until $t\simeq100\, t_a$ during which density fluctuations are slowly growing, and a second stage, during which density fluctuations abruptly increase. Although the density rms increases slightly faster than exponential in the second stage, especially for run~6, we estimated an average rate of $\gamma t_a=0.75$. Density growth saturates approximately at time $t\simeq180\,t_a$. In both run~5 and run~6, however, the effect of parametric decay is extremely weak and the cross helicity remains even larger than in the previous runs, around $\sigma\simeq0.96$ (cf. Fig.~\ref{sigma}).

\begin{figure}[htb]
 \includegraphics[width=0.53\textwidth]{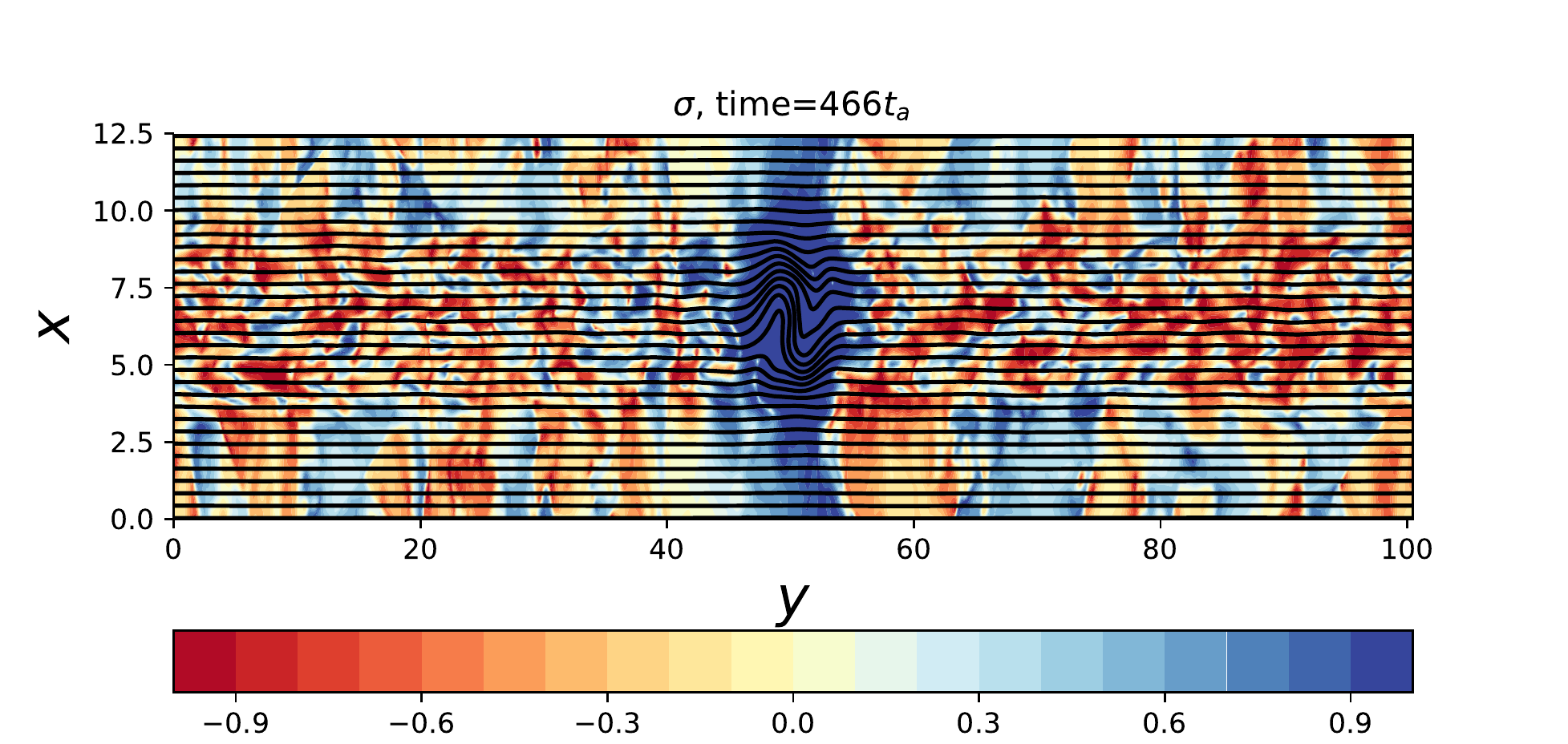}
  \includegraphics[width=0.53\textwidth]{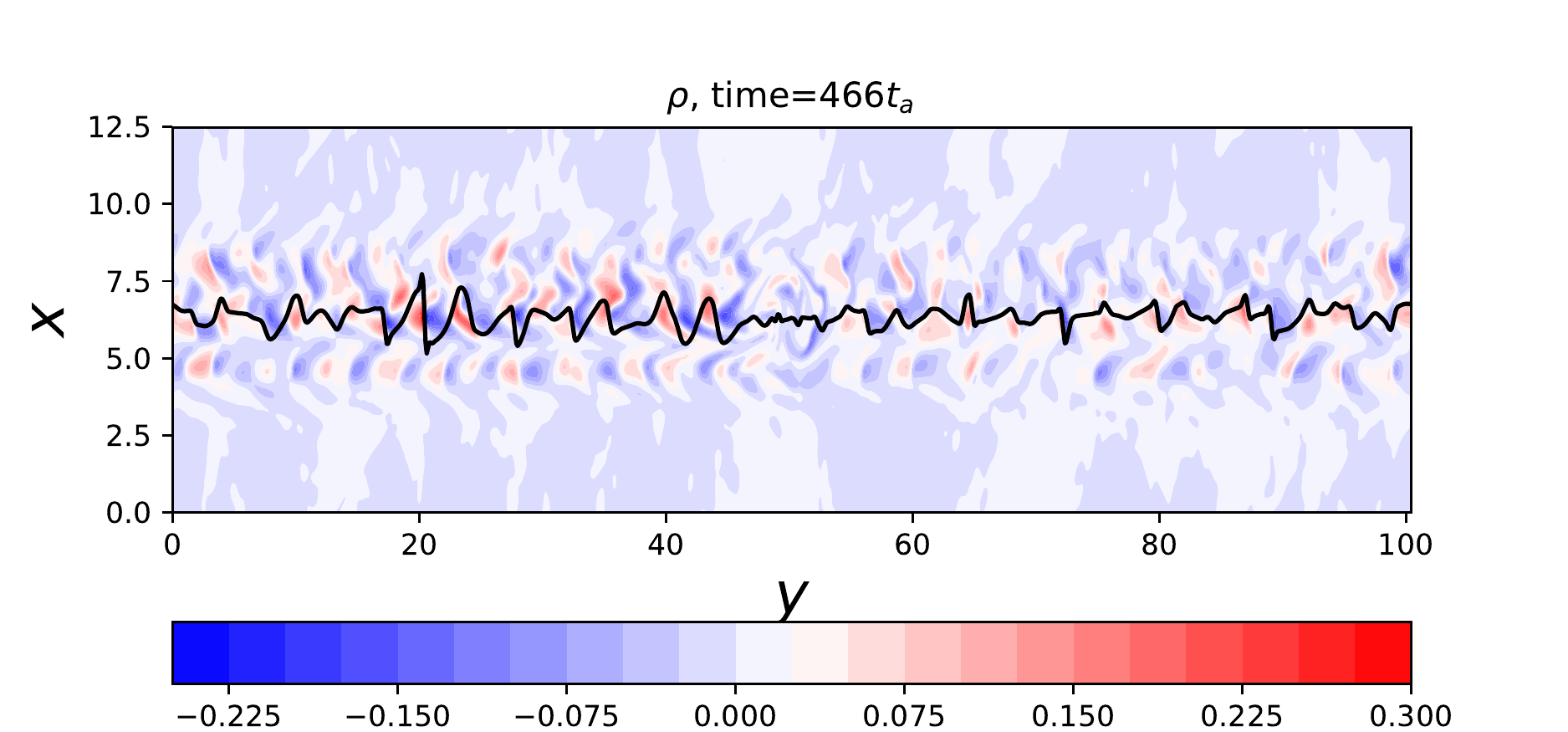}
\caption{Top panel: contour plot of the cross helicity $\sigma(x,y)$ (color coded) and magnetic field lines (black color) at time $t=466\,t_a$. Bottom panel:  contour plot of the density perturbation ($\rho-<\rho>$, where $<\cdot>$ indicates the spatial average) at time $t=466\,t_a$; the over-plotted black line represents the profile  of the density at the center of the switchback (run~5).}
\label{sigma5}
\end{figure}

\begin{figure}[htb]
  \includegraphics[width=0.5\textwidth]{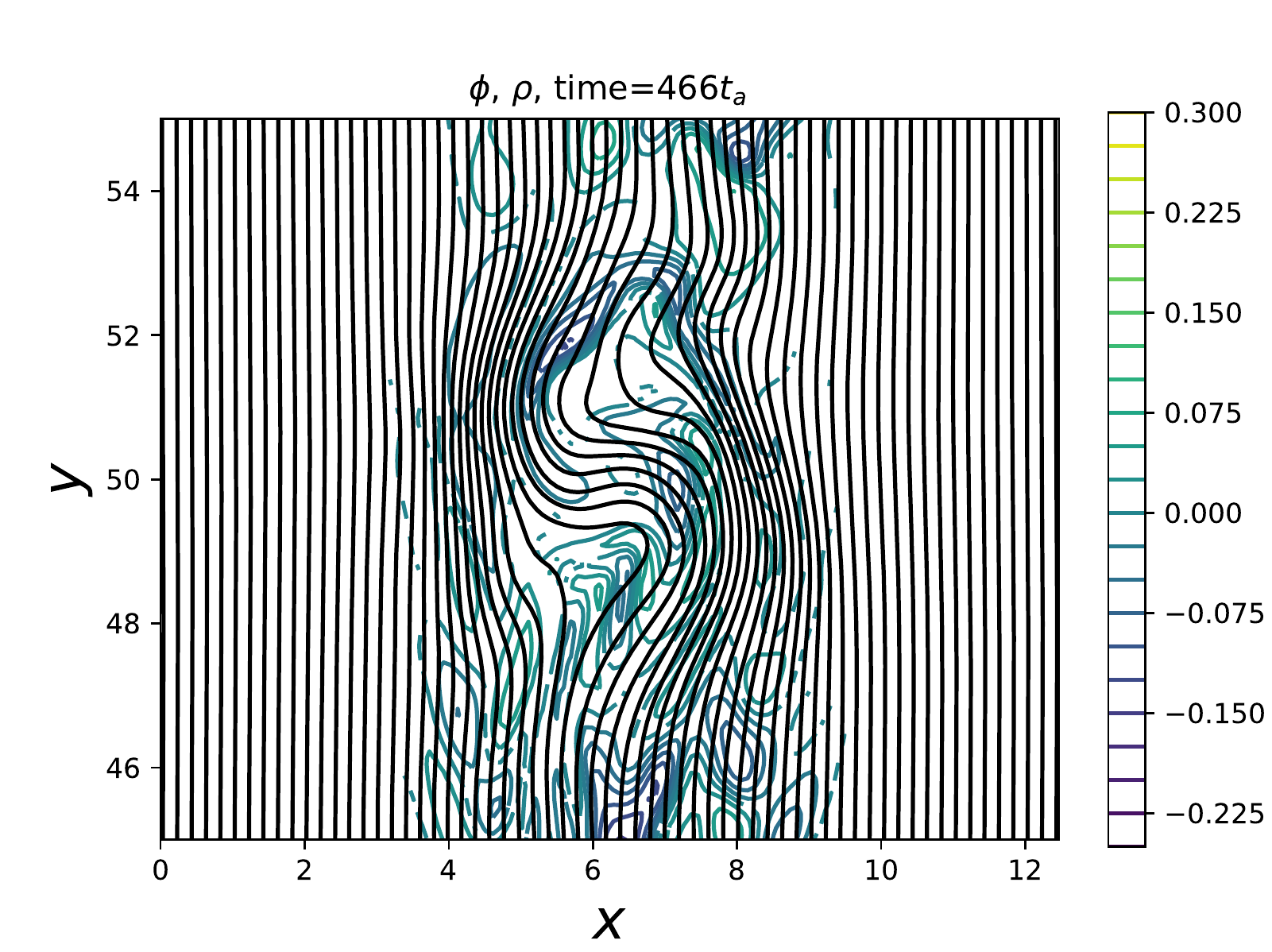}
\caption{Magnetic field lines (black) and contours of density fluctuations (color coded) at $t=466\,t_a$ (run 5). Only contour values corresponding to $|(\rho-\rho_0)/\rho_0|\geq10^{-3}$ are displayed. Only a portion of the simulation domain is shown.}
\label{evol_run5}
\end{figure}

In Fig.~\ref{sigma5}, top panel, we also show a contour of the cross helicity $\sigma(x,y)$ over the whole domain at $t=466\,t_a$ for run~5. As can be seen,  fluctuations propagating forwards and backwards, corresponding to positive and negative  cross-helicity, respectively (blue and red colors), are broadly excited outside  the switchback, but fields remain perfectly correlated, with $\sigma=+1$, at the switchback location. Density fluctuations are also excited, but as can be see from Fig.~\ref{sigma5}, bottom panel, they are depleted from inside the switchback. The switchback,  although reduced in strength, indeed persists beyond $t=466\,t_a$, as shown in Fig.~\ref{evol_run5}.  In this regard our work and results display some similarities but also differences from those of \citet{primavera_2019}, who also considered, starting from exact solutions to the MHD equations, non-planar (2D) fluctuations. In their work, \citet{primavera_2019} considered larger $\beta$ values, and their fluctuations were highly oblique  in the $(x,y)$ plane, filling space homogeneously.  In these conditions, they found that the parametric decay has a growth rate that increases with $\beta$, and that it leads to the formation of filamentary structures  characterized by density enhancements and backwards fluctuations which are strongly localized in a narrow band parallel to the propagation direction. In our case (see Fig.~\ref{sigma5}), density fluctuations and backward Alfv\'enic fluctuations are excited in a relatively narrow band as well, although in this case such a band clearly corresponds to the region traversed by the switchback.

\section{Discussion}

An ensemble of unidirectional Alfv\'en waves are an exact, attracting dynamical state within the framework of incompressible MHD~\citep{dvm}. If the total magnitude of the magnetic field is also constant, the solution is an exact one for compressible MHD~\citep{barnes_1974}, even in the presence of anisotropies, although it may become unstable in the presence of background fluctuations in the plasma quantities~\citep{anna2,anna3,anna4}. 

The new observations by PSP at $R\simeq 35.7\,R_\odot$ reveal the ubiquitous existence of Alfv\'enic fluctuations in the form of kinked and folded magnetic field lines. Understanding how long  they can survive in the solar wind is a first necessary step to understand their origin. In this regard our numerical simulations suggest that such a state can persist up to  hundreds of Alfv\'en times, in contrast to previous work where a similar kinked Alfv\'enic wave packet was considered, but in the case in which it was not in pressure equilibrium with its surrounding environment~\citep{Landi_2005}. 

If the background state is homogeneous, the process that eventually leads to a breakdown of the initial state is the well known parametric decay instability, that provides an upper bound to the lifetime of switchbacks, with increasing lifetime for increasing system size.
An order of magnitude estimate for the distance spanned by such fluctuations is therefore $\Delta R=\tau_{life}(V+v_a)$. In this regard our system size $L_y$ can be regarded as the  length scale traversed by a switchback before it undergoes a scattering event due to incoming perturbations. We can estimate $v_a$ and $V$ by taking the average of the Alfv\'en  and solar wind speed between $R=R_\odot$ and $R\simeq35\, R_\odot$, leading approximately to $\bar{v}_a\simeq 500$~km/s and $\bar V\simeq150$~km/s, respectively.  By considering a typical length of the switchback of $\ell_y\simeq5\times 10^4$~km we  obtain $t_a\simeq 100$~s and by assuming, in the worst case scenario, a small system size whose length scale is of the order of $L_y\simeq R_\odot$, which implies $L_y/\ell_y\simeq 10$, we fix $\tau_{life}\simeq200\,t_a$ (run1), yielding  $\Delta R\simeq 18\,R_\odot$. A larger distance of $\Delta R>43\,R_\odot$ will be spanned for larger values of $L_y/\ell_y$ (run 5 and 6).  

The above estimates must however be considered as indicative, for switchbacks propagate in a non-homogeneous, non-periodic expanding plasma and the effects of non-periodicity and underlying gradients must be taken into account simultaneously. 

Indeed, non-periodicity limits the growth time of a single parametric decay event to the time interval $\tau_i$ defined above. While in our periodic simulations these events have a waiting time $\Delta t\propto L_y$, in reality the latter is most probably an underestimate due to the stochastic arrival of incoherent fluctuations from the propagation direction, so that the lifetime of switchbacks may be even longer than what found in our simulations. On the other hand, although for $R/\ell_y\gg 1$ the  gradients in the expanding solar wind may be neglected in a local analysis,  expansion can have an integrated effect. Radial gradients, but also transverse gradients in the overall solar wind flow velocity as well as in the magnetic field and density, can lead to additional dynamical evolution that in part may lead to the disruption of the switchback, but in part may aid to maintain it with increasing distance from the sun. Indeed, not only expansion tends to inhibit the decay process~\citep{anna1} but also the presence of velocity jets associated with the switchback might provide a stabilizing mechanism, as discussed in the work of~\citet{Landi_2006} where it was shown that wakes and jets might build up a radial magnetic field inversion. 

In summary, different mechanisms may come into play at different distances from the sun and it is important to investigate further how such mechanisms may affect the disruption and re-generation of switchbacks during their propagation away from the sun, including the dependence of the decay process on the aspect ratio $\ell_y/\ell_x$ and amplitude of the switchback itself.

 \section{Summary}
We  have investigated via MHD simulations the evolution and stability properties of   a highly kinked  Alfv\'enic fluctuation that includes an inversion of the radial component of the magnetic field  and a total constant magnetic pressure, consistently with observations from PSP and earlier observations.  

We have shown that for a homogeneous, periodic system such Alfv\'enic wave packet/switchback can persist up to  several hundreds of Alfv\'en times, eventually decaying due to the onset of the parametric decay instability, that provides an upper bound to their lifetime. 

The lifetime of the switchback scales linearly with the system size due to the fact that it is a spatially localized wave-packet. For the largest system sizes considered here ($L_y/\ell_y=67 $ and $134$) the parametric decay is extremely slow and its effects are only weak (the initial switchback is reduced in strength but it remains folded for hundreds of Alfv\'en times). 

Using  average values of solar wind and Alfv\'en speeds, we have carried out an order of magnitude estimate of the distance  spanned by a switchback ($\Delta R$). We estimated that $\Delta R$ can range from $\Delta R\simeq18\,R_\odot$  (small system size) to $\Delta R> 43\, R_\odot$ (large system size), suggesting that switchbacks can be generated in the corona and propagate out to PSP distances, provided the wind is relatively calm, in the sense that there are not significant and frequent perturbations that may destroy them faster. 

The present study does not include the effects introduced by expansion and  other inhomogeneities, such as wind shears, that may alter the dynamics of switchbacks. We will need to investigate further those effects, as well as the role of the aspect ratio and amplitude of the switchback itself on the decay process.

\acknowledgements{Parker Solar probe was designed, built, and is now operated by the Johns Hopkins Applied Physics Laboratory as part of NASA's Living with a Star (LWS) program (contract NNN06AA01C).  The FIELDS and SWEAP experiments were developed and are operated under NASA contract NNN06AA01C. We also acknowledge NASA grant \#80NSS\-C18K1211 for supporting this research and the Texas Advanced Computing Center (TACC) at The University of Texas at Austin for providing HPC resources that have contributed to the research results reported within this paper. URL: http://www.tacc.utexas.edu.}

\end{document}